\documentclass[preprint,prd,nofootinbib,tightenlines,amsmath]{revtex4}
\usepackage{enumerate}
\usepackage{multirow}

\usepackage{epsfig}
\usepackage{graphics,color}      
\usepackage{subcaption}
\usepackage{verbatim}
\usepackage{amsmath}   
\usepackage[section]{placeins}
\usepackage[normalem]{ulem} 
\usepackage{soul}
\usepackage{mathrsfs}

\usepackage[colorlinks=true,
            linkcolor=red,
            urlcolor=blue,
            citecolor=blue]{hyperref}
\usepackage{xcolor}
\definecolor{orange}{rgb}{1,0.5,0}
\definecolor{amethyst}{rgb}{0.6, 0.4, 0.8}
\definecolor{antiquefuchsia}{rgb}{0.57, 0.36, 0.51}
\definecolor{brightlavender}{rgb}{0.75, 0.58, 0.89}

\catcode`\@=11
\def\sla@#1#2#3#4#5{{%
 \setbox\z@\hbox{$\m@th#4#5$}%
 \setbox\tw@\hbox{$\m@th#4#1$}%
 \dimen4\wd\ifdim\wd\z@<\wd\tw@\tw@\else\z@\fi
 \dimen@\ht\tw@
 \advance\dimen@-\dp\tw@ \advance\dimen@-\ht\z@
 \advance\dimen@\dp\z@
 \divide\dimen@\tw@ \advance\dimen@-#3\ht\tw@
 \advance\dimen@-#3\dp\tw@ \dimen@ii#2\wd\z@
 \raise-\dimen@\hbox to\dimen4{%
 \hss\kern\dimen@ii\box\tw@\kern-\dimen@ii\hss}%
 \llap{\hbox to\dimen4{\hss\box\z@\hss}}}}

\def\declareslashed#1#2#3#4#5{%
 \expandafter\def\csname sla@\string#5\endcsname{%
#1{\mathpalette{\sla@{#2}{#3}{#4}}{#5}}}}
 \catcode`\@=12
\declareslashed{}{/}{.08}{0}{D}
 \declareslashed{}{/}{.1}{0}{A}
 \declareslashed{}{/}{0}{-.05}{k}
 \declareslashed{}{/}{.1}{0}{\partial}
 \declareslashed{}{\not}{-.6}{0}{f}

\def\lsim{\mathrel {\vcenter {\baselineskip 0pt \kern 0pt
    \hbox{$<$} \kern 0pt \hbox{$\sim$} }}}
\def\gsim{\mathrel {\vcenter {\baselineskip 0pt \kern 0pt
    \hbox{$>$} \kern 0pt \hbox{$\sim$} }}}
    
\newcommand{\gev}{~\text{GeV}}

\begin{document}

\baselineskip=15pt

\title{Scrutinizing a massless dark photon:\\
basis independence}

\author{Jun-Xing Pan${}^{1}$\footnote{panjunxing2007@163.com}}
\author{Min He${}^{2}$\footnote{hemind@sjtu.edu.cn}}
\author{Xiao-Gang He${}^{2,1,3,4}$\footnote{hexg@phys.ntu.edu.tw}}
\author{Gang Li$^{3}$\footnote{gangli@phys.ntu.edu.tw}}

\affiliation{${}^{1}$School of Physics and Information Engineering, Shanxi Normal University, Linfen 041004, China}
\affiliation{${}^{2}$Tsung-Dao Lee Institute, and School of Physics and Astronomy, Shanghai Jiao Tong University, Shanghai 200240, China}
\affiliation{${}^{3}$Department of Physics, National Taiwan University, Taipei 10617, Taiwan}
\affiliation{${}^{4}$Physics Division, National Center for Theoretical Sciences, Hsinchu 30013, Taiwan}

\date{\today}

\vskip 1cm
\begin{abstract}

A new $U(1)_X$ gauge boson field $X$  can have renormalizable kinetic mixing with the standard model (SM) $U(1)_Y$ gauge boson field $Y$. This mixing induces interactions of $X$ with SM particles even though $X$ starts as a dark photon without such interactions. If the $U(1)_X$ is not broken, both the dark photon field $X$ and the photon field $A$ are massless. One cannot determine which one of them is the physical dark photon or the photon by just looking at kinetic terms in the Lagrangian. We revisit this issue and show explicitly that when all contributions are included, all physical processes do not depend on which basis is used  and the kinetic mixing effects do not show up in electromagnetic and weak interactions if only SM particles are involved in the calculations. On the other hand, the kinetic mixing provides a portal for probing the dark sector beyond the SM. We update constraints on the millicharged dark sector particles from the Lamb shift and lepton $g-2$ measurements.

\end{abstract}

\pacs{PACS numbers: }

\maketitle
    

\newpage

\section{Introduction}
A new gauge symmetry $U(1)_X$ with a gauge boson field $X$ can mix with 
the $U(1)_Y$ gauge boson field $Y$ of $U(1)_Y$ in the standard model (SM) through a renormalizable kinetic mixing operator $X_{\mu\nu} Y^{\mu\nu}$ formed by the field strengths, $F_{\mu\nu} = \partial_\mu F_\nu - \partial_\nu F_\mu$, with $F=X,Y$~\cite{Holdom:1985ag,Foot:1991kb,Okun:1982xi,Galison:1983pa}. If the SM particles are all uncharged under the $U(1)_X$, it is expected to have no interaction with SM particles. In this case $X$ is dubbed as a dark photon field. However, the kinetic mixing term can induce interactions between $X$ and the SM particles. This has many interesting consequences in low energy and high energy phenomena from particle physics, astrophysics to cosmology perspectives. Dark photon has been searched for in a number of different contexts experimentally~\cite{Essig:2013lka,Alexander:2016aln}.

If the dark photon field $X$ receives a finite mass, one can easily identify the physical dark photon and photon after the fields are redefined to have the canonical form for the gauge bosons, in which the kinetic terms are diagonal. 
However, for the case that the dark photon is trivially massless,
the situation is different. If one just looks at the kinetic terms of $X$ and $Y$, the canonical form is invariant under any orthogonal transformation, then one cannot tell any difference before and after the transformation. Therefore, which combination of $X$ and $Y$ in the canonical form corresponds to the physical photon or dark photon cannot be determined~\cite{delAguila:1995rb}.

Phenomenology of a massless dark photon has drawn a vast of attention~\cite{Dobrescu:2004wz,Fabbrichesi:2017zsc,Fabbrichesi:2017vma,Gabrielli:2014oya,
Biswas:2016jsh,Daido:2018dmu,Huang:2018mkk,Ackerman:mha}. One needs to be clear about how the massless dark photon interacts with SM particles to have correct interpretations of the results. The interactions of photon, $Z$ boson and dark photon fields to the SM currents must be consistently defined to pin down the massless dark photon itself. 
We find that two commonly used ways to remove the mixing term are actually related through an orthogonal transformation. But the angle that describes the general orthogonal transformation does not affect how the massless dark photon and photon interact with SM particles. 
We show that effects of the kinetic mixing does not leave traces in the electromagnetic (EM) and weak interactions involving only SM particles (in loops or external states), such as $g-2$ of a charged lepton and in the processes Higgs decays into two photons, and $Z$ boson decays into SM particles. To detect massless dark photon effects, information about dark current needs to be known in some way, either in the form of missing energy (small ionization energy loss) with dark current at tree level or in SM measurements with dark current at loop level.

This paper is organized as follows. In Sec.~\ref{sec:formalism}, we will show the interactions of photon and dark photon fields to the SM currents and dark current in a general basis. In Sec.~\ref{sec:main}, we will discuss the physical effects of a massless dark photon. Sec.~\ref{sec:summary} summarizes our results.

\section{Eliminating kinetic mixing for a massless dark photon}
\label{sec:formalism}

With the kinetic mixing, the kinetic terms of $X$ and $Y$ and their interactions with other particles can be written as 
\begin{eqnarray}
\mathcal{L} = -{1\over 4} X_{\mu\nu} X^{\mu\nu} - {\sigma \over 2} X_{\mu\nu} Y^{\mu\nu} - {1\over 4} Y_{\mu\nu} Y^{\mu\nu} + j^\mu_Y Y_\mu + j^\mu_X X_\mu\;.
\end{eqnarray}
Here $j^\mu_X$ and $j^\mu_Y$ denote interaction currents of gauge fields $X$ and $Y$, respectively.

To write the above Lagrangian in the canonical form one needs to diagonalize the kinetic terms of $X$ and $Y$.
Let us consider two commonly used ways of removing the mixing, namely, a)~\cite{Holdom:1985ag,Dobrescu:2004wz} the mixing term  is removed in such a way that dark photon $\hat X$ in the canonical form does not couple to hyper-charge current $j^\mu_Y$~\footnote{In the literature case a) is widely used not only for a massless dark photon but also for a very light one~\cite{Ahlers:2007rd,Redondo:2008aa,Danilov:2018bks}, which is sometimes called ``paraphoton''~\cite{Okun:1982xi,Holdom:1985ag}.}, and b)~\cite{Foot:1991kb,Babu:1997st} the hyper-charge field in the 
canonical form $\hat Y^\prime$ does not couple to  dark current $j^\mu_X$ produced by some dark particles with $U(1)_X$ charges, which is widely used in the studies of a massive dark photon or $Z^\prime$~\cite{Foot:1991kb,He:2017ord,He:2017zzr,Babu:1997st,Curtin:2014cca}. 
For the cases a) and b), making the Lagrangian in the canonical form will be
\begin{eqnarray}
Case\;\; a):&& \mathcal{L}_a = -{1\over 4} \hat X_{\mu\nu} \hat X^{ \mu\nu} -{1\over 4} \hat Y_{\mu\nu} \hat Y^{ \mu\nu}+ j^\mu_Y{1\over \sqrt{1-\sigma^2}} \hat Y_\mu  + j^\mu_X  ( \hat X_\mu -{\sigma \over \sqrt{1-\sigma^2}} \hat Y_\mu)\;,\nonumber\\
&&\hat Y_\mu = \sqrt{1-\sigma^2} Y_\mu\;,\;\; \hat  X_\mu = \sigma Y_\mu + X_\mu\;,\nonumber\\
Case\;\; b):&& \mathcal{L}_b = -{1\over 4} \hat X^\prime_{\mu\nu} \hat X^{\prime \mu\nu} -{1\over 4} \hat Y^\prime_{\mu\nu} \hat Y^{\prime \mu\nu}+ j^\mu_Y( \hat Y^\prime_\mu -{\sigma\over \sqrt{1-\sigma^2}} \hat X^\prime_\mu) + j^\mu_X {1\over \sqrt{1-\sigma^2}} \hat X^\prime_\mu\;,\nonumber\\
&&\hat Y^\prime_\mu = Y_\mu + \sigma X_\mu\;,\;\;\hat X^\prime_\mu = \sqrt{1-\sigma^2} X_\mu\;.
\end{eqnarray}

After electroweak symmetry breaking (EWSB), the hyper-charge field $Y$ and the neutral component of the $SU(2)_L$ gauge field $W^3$ can be written in the combinations of the ordinary photon field $A$ and the $Z$ field as follows
\begin{eqnarray}
Y_\mu = c_W A_\mu -s_W Z_\mu\;,\;\; W^3_\mu = s_W A_\mu + c_W Z_\mu\;,
\end{eqnarray}
where $c_W \equiv \cos\theta_W$ and $s_W \equiv \sin\theta_W$ with $\theta_W$ being the weak mixing angle. Meanwhile, the $Z$ field receives a mass $m_Z$.

The general Lagrangian that describes $A$, $Z$ and $X$ fields kinetic energy, and their interactions with the electromagnetic (EM) current $j^\mu_{em}$, neutral $Z$-boson current $j^\mu_Z$ and dark current $j^\mu_X$ are given by
\begin{eqnarray}
\mathcal{L} = &&-{1\over 4} X_{\mu\nu} X^{\mu\nu} - {1\over 4} A_{\mu\nu} A^{\mu\nu} - {1\over 4} Z_{\mu\nu} Z^{\mu\nu} 
-{1\over 2} {\sigma c_W}X_{\mu\nu} A^{\mu\nu} + {1\over 2} {\sigma s_W} X_{\mu\nu} Z^{\mu\nu}\nonumber\\
&& + j^\mu_{em} A_\mu + j^\mu_Z Z_\mu + j^\mu_X X_\mu +{1\over 2}m^2_Z Z_\mu Z^\mu\;,
\end{eqnarray}
where the $Z$ boson mass term is included. 

The dark photon may be also massive. There are two popular ways of generating dark photon mass giving rise to different phenomenology. One of them is the ``Higgs mechanism'', in which the $U(1)_X$ is broken by the vacuum expectation value (vev) of a SM singlet, which is charged under $U(1)_X$. In this case, the mixing of Higgs doublet and the Higgs singlet offers the possibility of searching for dark photon at colliders in Higgs decays~\cite{Curtin:2014cca}. The other is the ``Stueckelberg mechanism''~\cite{Kors:2004dx,Stueckelberg:1900zz} in which an axionic scalar was introduced to allow a mass for $X$ without breaking $U(1)_X$. An interesting application of this mechanism to a gauged B-L symmetry has been discussed in Ref.~\cite{Heeck:2014zfa}. In our later discussion our concern is whether the dark photon has a mass or not, and therefore we only need to discussion the effect of a  mass term $(1/2)m^2_X X_\mu X^\mu$ in the above equation. The $W^\pm$ fields and their mass, due to electroweak symmetry breaking of the SM, are not affected.

The requirements for cases a) and b) can be equivalently expressed as no dark photon interaction with $j^\mu_{em}$ and no photon interaction 
with $j^\mu_X$, respectively. These two cases can be achieved by defining
\begin{eqnarray}
Case\;\; a):&&\left (\begin{array}{l}
A\\
Z\\
X
\end{array}
\right )
= \left ( \begin{array}{ccc}
{1\over \sqrt{1-\sigma^2c^2_W}}& {- \sigma^2s_{W}c_{W}\over \sqrt{1-\sigma^2}\sqrt{1-\sigma^2c^2_W}}&0\\
0&{\sqrt{1-\sigma^2c^2_W}\over \sqrt{1-\sigma^2}}&0\\
{-\sigma c_W \over \sqrt{1-\sigma^2c^2_W}}&{\sigma s_W\over \sqrt{1-\sigma^2}\sqrt{1-\sigma^2c^2_W}}&1
\end{array}
\right )
\left (\begin{array}{l}
\tilde A\\
\tilde Z\\
\tilde X
\end{array}
\right )\;,\nonumber\\
\\\nonumber\\
Case\;\;b):&&\left (\begin{array}{l}
A\\
Z\\
X
\end{array}
\right )
=\left ( \begin{array}{ccc}
1&{-\sigma^2s_W c_W\over \sqrt{1-\sigma^2}\sqrt{1-\sigma^2 c^2_W}}&{-\sigma c_W\over \sqrt{1-\sigma^2 c^2_W}}\\
0&{\sqrt{1-\sigma^2c^2_W}\over \sqrt{1-\sigma^2 }}&0\\
0&{\sigma s_W\over \sqrt{1-\sigma^2}\sqrt{1-\sigma^2c^2_W}}&{1\over \sqrt{1-\sigma^2c^2_W}}
\end{array}
\right )
\left (\begin{array}{l}
\tilde A^\prime\\
\tilde Z^\prime\\
\tilde X^\prime
\end{array}
\right )\;,\nonumber
\end{eqnarray}
to obtain the Lagrangian  in the case of $m_X =0$,
\begin{eqnarray}
\label{eq:case_a_azx}
\mathcal{L}_a = &&-{1\over 4} \tilde X_{\mu\nu} \tilde X^{\mu\nu} - {1\over 4} \tilde A_{\mu\nu} \tilde A^{\mu\nu} - {1\over 4} \tilde Z_{\mu\nu} \tilde Z^{\mu\nu} +{1\over 2} m^2_Z {1-\sigma^2c^2_W\over 1-\sigma^2} \tilde Z_\mu \tilde Z^\mu
\nonumber\\
&& + j^\mu_{em} ({1\over \sqrt{1-\sigma^2c^2_W}}\tilde A_\mu   - {\sigma^2s_W c_W\over \sqrt{1-\sigma^2} \sqrt{1-\sigma^2c^2_W}}\tilde Z_\mu ) + j^\mu_Z ( {\sqrt{1-\sigma^2 c^2_W}\over \sqrt{1-\sigma^2}} \tilde Z_\mu ) \\
&&+ j^\mu_X( {-\sigma c_W\over  \sqrt{1-\sigma^2c^2_W}}\tilde A_\mu + {\sigma s_W\over \sqrt{1-\sigma^2}\sqrt{1-\sigma^2 c^2_W}}\tilde Z_\mu + \tilde X_\mu)\;,\nonumber\\
\label{eq:case_b_azx}
\mathcal{L}_b = &&-{1\over 4} \tilde X^\prime_{\mu\nu} \tilde X^{\prime \mu\nu} - {1\over 4} \tilde A^\prime_{\mu\nu} \tilde A^{\prime \mu\nu} - {1\over 4} \tilde Z^\prime_{\mu\nu} \tilde Z^{\prime \mu\nu} +{1\over 2} m^2_Z {1-\sigma^2c^2_W\over 1-\sigma^2} \tilde Z^\prime_\mu \tilde Z^{\prime \mu}
\nonumber\\
&& + j^\mu_{em} (\tilde A^\prime_\mu   - {\sigma^2s_W c_W\over \sqrt{1-\sigma^2} \sqrt{1-\sigma^2c^2_W}}\tilde Z^\prime_\mu -{\sigma c_W\over  \sqrt{1-\sigma^2c^2_W}}\tilde X^\prime_\mu )\\
&& + j^\mu_Z ( {\sqrt{1-\sigma^2 c^2_W}\over \sqrt{1-\sigma^2}} \tilde Z^\prime_\mu )+ j^\mu_X(  {\sigma s_W\over \sqrt{1-\sigma^2}\sqrt{1-\sigma^2 c^2_W}}\tilde Z^\prime_\mu +{1\over \sqrt{1-\sigma^2 c^2_W}} \tilde X^\prime_\mu )\;.\nonumber
\end{eqnarray}
We clearly see that the properties for case a) and case b) are explicit.  In both cases the $Z$ boson mass is shifted as $m_Z^2 \to m_Z^2(1+z)$ with $z=\sigma^2s_W^2/(1-\sigma^2)$. Note that in the above two ways of removing the kinetic mixing term, the $Z$ boson interactions are the same in form. 

The dark photon fields in the above are $\tilde X$ and $\tilde X^\prime$, respectively. It has been argued using Eq.~\eqref{eq:case_a_azx} that dark photon does not interact with SM particles at the tree-level~\cite{Dobrescu:2004wz,Fabbrichesi:2017zsc,Fabbrichesi:2017vma,Gabrielli:2014oya}. But if one uses Eq.~\eqref{eq:case_b_azx}, the dark photon does interact with SM particles at the tree-level. The statements are in conflict with each other. This conflict lies in the definition for a dark photon. 

If one just looks at the first two kinetic terms in Eqs.~\eqref{eq:case_a_azx}~\eqref{eq:case_b_azx}, they are the same in form and invariant under an orthogonal transformation of 
$\tilde X$ and $\tilde A$, or $\tilde X^\prime$ and $\tilde A^\prime$.  In fact, there are related by
\begin{eqnarray}
\left ( \begin{array}{c}
\tilde A^\prime\\
\tilde X^\prime
\end{array}
\right ) 
= \left (\begin{array}{cc}
\sqrt{1-\sigma^2c^2_W}&\sigma c_W\\
&\\
-\sigma c_W&\sqrt{1-\sigma^2c^2_W}
\end{array}
\right )
\left (\begin{array}{c}
\tilde A\\
\tilde X
\end{array}
\right )\;. \label{transform}
\end{eqnarray}

But for the case with $m_X \neq 0$, the situation is different. One can completely determine the physical states among $A$, $Z$ and $X$. 
With $m_X \neq 0$, we need to add a mass term $(1/2)m^2_X X_\mu X^\mu$ to the Lagrangian. In cases a) and b), they have the following forms
\begin{eqnarray}
Case\;\; a):&&{1\over 2} m^2_X ( {-\sigma c_W\over  \sqrt{1-\sigma^2c^2_W}}\tilde A_\mu + {\sigma s_W\over \sqrt{1-\sigma^2}\sqrt{1-\sigma^2 c^2_W}}\tilde Z_\mu + \tilde X_\mu)^2\;,\nonumber\\
Case\;\;b):&&{1\over 2} m^2_X (  {\sigma s_W\over \sqrt{1-\sigma^2}\sqrt{1-\sigma^2 c^2_W}}\tilde Z^\prime_\mu +{1\over \sqrt{1-\sigma^2 c^2_W}} \tilde X^\prime_\mu )^2\;.
\end{eqnarray}

To identify the physical photon, we find that the fields defined in case b) is more convenient to use since the field $\tilde A^\prime$ is
already the physical massless photon field $A^m$ without further mass diangonalization. To obtain physical $Z^m$ and $X^m$, in case b), one needs to diagonalize the mass matrix in $(\tilde Z^\prime, \tilde X^\prime)$ basis,
\begin{eqnarray}
\left ( \begin{array}{cc}
{m^2_Z (1-\sigma^2 c^2_W)^2 + m^2_X \sigma^2 s^2_W\over (1-\sigma^2)(1-\sigma^2 c^2_W)}&{m^2_X \sigma s_W\over \sqrt{1-\sigma^2}(1-\sigma^2c_W^2)}\\
{m^2_X \sigma s_W\over \sqrt{1-\sigma^2}(1-\sigma^2c_W^2)}& {m^2_X\over 1-\sigma^2c^2_W}
\end{array}
\right )\;,
\end{eqnarray}
to obtain the mass eigenstates
\begin{eqnarray}
&&\left (\begin{array}{c}
Z^m\\
X^m
\end{array}
\right )
=
\left (\begin{array}{cc}
\cos\theta&\sin\theta\\
-\sin\theta&\cos\theta
\end{array}
\right )
\left (\begin{array}{c}
\tilde Z^\prime\\
\tilde X^\prime
\end{array}
\right )\;,
\end{eqnarray}
with
\begin{eqnarray}
\tan(2\theta) = {2m^2_X \sigma s_W\sqrt{1-\sigma^2} \over m^2_Z (1-\sigma^2 c^2_W)^2 - m^2_X [1- \sigma^2(1+s^2_W)]}\;.
\end{eqnarray}

The interactions of physical photon, $Z$ boson and dark photon can be determined accordingly without ambiguities. Expressing $\tilde A^\prime$, $\tilde Z^\prime$ and $\tilde X^\prime$ in terms of $A^m$, $Z^m$ and $X^m$, one also obtain physical gauge boson interactions with SM and dark sector particles. A consistent treatment for case a) will lead to the same final results.

Let us come back to the situation with $m_X =0$ and discuss whether one can determine what the physical photon and massless dark photon are.
To this end we use a most general basis $(\bar A^\prime, \bar X^\prime, \bar{Z}^\prime)$ based on case b)
\begin{eqnarray}
\left ( \begin{array}{c}
\tilde A^\prime\\
\tilde X^\prime
\end{array}
\right ) 
= \left (\begin{array}{cc}
c_\beta &s_\beta\\
-s_\beta &c_\beta
\end{array}
\right )
\left (\begin{array}{c}
\bar A^\prime\\
\bar X^\prime
\end{array}
\right )\;,\;\; \tilde Z^\prime = \bar Z^\prime \;, \label{transform1}
\end{eqnarray}
where $c_\beta \equiv \cos\beta$ and $s_\beta \equiv \sin\beta$. For $s_\beta = \sigma c_W$, $\bar A^\prime = \tilde A$ and $\bar X^\prime = \tilde X$ as compared with Eq.~\eqref{transform}. For $\beta$ spanning from 0 to $2\pi$, all possible ways of removing the kinetic mixing to have a canonical form of $A$, $Z$ and $X$ fields can be covered. We have the following Lagrangian for the most general form for interactions for $\bar{A}^\prime$, $\bar{Z}^\prime$ and $\bar{X}^\prime$
\begin{eqnarray}
\label{eq:Lb-bar}
\mathcal{L}_{\bar b} = &&-{1\over 4} \bar X^\prime_{\mu\nu} \bar X^{\prime \mu\nu} - {1\over 4} \bar A^\prime_{\mu\nu} \bar A^{\prime \mu\nu} - {1\over 4} \bar Z^\prime_{\mu\nu} \bar Z^{\prime \mu\nu} +{1\over 2} m^2_Z {1-\sigma^2c^2_W\over 1-\sigma^2} \bar Z^\prime_\mu \bar Z^{ \prime \mu}
\nonumber\\
&& +\left ( (c_\beta   + {\sigma c_W\over  \sqrt{1-\sigma^2c^2_W}} s_\beta)j^\mu_{em}   -  s_\beta \dfrac{1}{\sqrt{1-\sigma^2 c_W^2}} j^\mu_X \right ) \bar A^\prime_\mu \nonumber\\
&& +\left ( {\sqrt{1-\sigma^2 c^2_W}\over \sqrt{1-\sigma^2}}  j^\mu_Z   - {\sigma^2s_W c_W\over \sqrt{1-\sigma^2} \sqrt{1-\sigma^2c^2_W}}  j^\mu_{em}  +{\sigma s_W\over \sqrt{1-\sigma^2}\sqrt{1-\sigma^2 c^2_W}}  j^\mu_X  \right )\bar Z^\prime_\mu\nonumber\\
&&  + \left ( {1\over \sqrt{1-\sigma^2 c^2_W}}c_\beta j^\mu_X   +  (s_\beta  -{\sigma c_W\over  \sqrt{1-\sigma^2c^2_W}}c_\beta) j^\mu_{em} \right ) \bar X^\prime_\mu \;.
\end{eqnarray}

Note that in the above $\bar A^\prime$ and $\bar X^\prime$ are not what to be identified as physical photon and dark photon. In the presence of dark current, the physical photon $\gamma$ and dark photon $\gamma_D$ should be the fields which respond to $j^\mu_{em}$ and $j^\mu_{X}$ to produce signal, that is, the components in $\bar A^\prime$ and $\bar X^\prime$ to $j^\mu_{em}$ and $j^\mu_{X}$, respectively.  In next section, we will use $A$ and $X$ to stand for the fields $\bar{A}^{\prime}$ and $\bar{X}^{\prime}$ for convenience.

\section{Physical effects of a massless dark photon }
\label{sec:main}

Let us first study how EM interaction is affected by the kinetic mixing of a massless dark photon to the order of approximation where no dark sector particles, except the dark photon, are involved in either loop or initial and final states. A well-motivated observable is the anomalous magnetic dipole moment $g-2$ of fermion. There is a longstanding discrepancy between the experimental value and the SM prediction of the anomalous magnetic moment of the muon, $a_{\mu}=(g-2)_{\mu}/2$~\cite{Tanabashi:2018zz}.
A lot of theoretical efforts have been made to explain this anomaly, see Refs.~\cite{Baek:2001kca,Pospelov:2008zw} for the ``solutions'' with a massive dark photon. 
 
\begin{figure}
\includegraphics[width=0.2\textwidth]{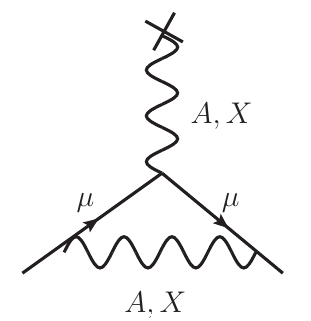}
\caption{The Feynman diagram that contributes to muon $g-2$ at one-loop level. The cross vertex denotes the external source.}
\label{fig:g-2}
\end{figure}

Supposing that the external field is the SM photon $\gamma$, the contribution of massless dark photon field $X$ to the muon $g-2$ can be easily obtained from rescaling the one-loop EM correction by a factor $R_X$, that is $R_X {\alpha}/{(2\pi)}$ with
\begin{eqnarray}
\label{eq:muon_g-2}
\quad R_X=\left(  s_\beta - {\sigma c_W \over \sqrt{1-\sigma^2 c^2_W}}c_\beta \right)^2, \label{R}
\end{eqnarray} 
where $\alpha = e^2/4\pi$. Explaining $\Delta a_{\mu}$ with $R_X {\alpha}/{(2\pi)}$ seemingly indicates physical effect of the massless dark photon depending on the artificial rotation angle $\beta$.

However, since both photon and dark photon fields are massless, their contributions to muon $g-2$ should be included consistently. Apart from the massless dark photon field $X$, the photon field $A$ in the loop should also be taken into account. Adding them up, we obtain
\begin{eqnarray}
(R_X+R_A) \dfrac{\alpha}{2\pi} = {1\over 1-\sigma^2 c_W^2} \dfrac{\alpha}{2\pi} =\dfrac{\bar \alpha}{2\pi},
\end{eqnarray} 
where
\begin{eqnarray}
\quad R_A=\left(  c_\beta + {\sigma c_W \over \sqrt{1-\sigma^2 c^2_W}}s_\beta \right)^2 \label{R_A}
\end{eqnarray}
and ${\bar \alpha} = {\bar e}^2/{4\pi} = \alpha / (1-\sigma^2 c_W^2)$ with the redefinition of the electric charge $\bar e = e/\sqrt{1-\sigma^2 c_W^2}$. This amounts to redefine $\bar j^\mu_{em} = (\bar e/e)j^\mu_{em}$. The effect of kinetic mixing term can therefore to be absorbed into redefinition of the electric charge that is independent of the angle $\beta$. For convenience we will also redefine the dark gauge coupling $g_X$ to be $\bar g_X = g_X/\sqrt{1-\sigma^2c^2_W}$. The experimental extraction of muon $g-2$ cannot distinguish the external fields $A$ and $X$, which couple to the external source and muon differently from the SM photon. However, the effect can also be absorbed into the electric charge redefinition. The complete one-loop correction to muon $g-2$ is shown in Fig.~\ref{fig:g-2}.

The charge redefinition of can also be seen in the Compton scattering, which is used to measure the fine structure constant experimentally~\cite{Parker191}. Moreover, in the general basis, we find that only with the summation of the cross sections of $Ae^-\to Ae^-$, $Ae^-\to Xe^-$, $Xe^-\to Ae^-$ and $Xe^-\to Xe^-$, the total cross section in the Compton scattering is basis independent and equal to the one calculated in case a).

In the above we have seen that the observable effects of EM interaction is independent of $\beta$ in the example of muon $g-2$ and no beyond SM effects show up. From Eq.~\eqref{eq:Lb-bar}, the $Z$ boson interaction is already independent of $\beta$. One can ask whether any physical effects induced by the kinetic mixing show up in the weak interaction involving only SM particles and the massless dark photon.

In the minimal massless dark model we considered, there is no modification of $W$ boson interactions. Thus no new effects will show up in weak interactions involving the $W$  
bosons. The mass of $W^\pm$ and the charged current $j^{\pm \mu}_W$
are not affected by the field redefinition as discussed in section~\ref{sec:formalism}, we have $\bar m^2_W = m^2_W=c_W^2 m_Z^2$, and the charged current $\bar j_W^{\pm\mu} =j_W^{\pm \mu}$. But the $Z$ boson mass is modified as $\overline {m}^2_Z = m^2_Z(1+z)$. Closely following Ref.~\cite{Burgess:1993vc}, we choose the three input parameters as fine structure constant, Fermi constant, and the $Z$ boson mass and calculate the other observables. The weak mixing angle can be determined in terms of these three input parameters: $\bar{s}_W\bar{c}_W=\left[\pi\bar{\alpha}/\left(\sqrt{2}G_F \bar{m}_Z^2\right)\right]^{1/2}$~\cite{Bardin:1997xq,Langacker:2010zza}, which shifts as $\bar{s}_W^2=s_W^2\left[ 1+c_W^2/(c_W^2-s_W^2)(\mathcal{A}-\mathcal{C}) \right]$ with $\mathcal{A} \equiv -\sigma^2 c_W^2/(1-\sigma^2 c_W^2) $ and $\mathcal{C} \equiv -\sigma^2 s_W^2/(1-\sigma^2)$. Besides, we have the following redefined  currents
\begin{eqnarray}
&&\bar j^\mu_{em} = \bar e Q_f \bar f \gamma^\mu f\;,\;\;\bar j^{+\mu}_W  = {\bar g_W\over 2 \sqrt{2}} \bar f^u\gamma^\mu(1-\gamma_5) f^d\;,\;\;\bar j^\mu_Z = {\bar g_Z\over 2}
\bar f \gamma^\mu (\bar{g}_{V}^{f} -\bar{g}_{A}^{f}\gamma^5) f\;,
\end{eqnarray}
where $f^{u}$ and $f^d$ indicate the upper and lower components of the left handed fermion doublets, respectively. The gauge couplings $\bar g_W=2 (\sqrt{2}G_F \bar{m}^2_W )^{1/2}$, $\bar g_Z=2 (\sqrt{2}G_F \bar{m}^2_Z )^{1/2}$. The axial-vector and vector couplings of $Z$ boson are given by
\begin{eqnarray}
\overline{ g}^f_A =   I^3_f\;,\;\;\overline{ g}^f_V =   (I^3_f - 2 Q_f\overline{s}^2_*)
\end{eqnarray}
with $\overline{ s}^2_* = s^2_W [1+{\sigma^2  c^2_W /(1-\sigma^2 c^2_W})]$.
Here $f$ indicates SM fermion which has the $SU(2)_L$ isospin $I^3_f$ to be $1/2$ and $-1/2$ for the upper and down doublet components.

The $\rho$ parameter, which measures the ratio of low-energy neutral- and charged-current amplitudes, can be expressed as~\cite{Barger:1987nn,Burgess:1993vc}
\begin{eqnarray}
\rho \equiv {g^2_Z/\bar m^2_Z \over g^2/\bar m^2_W}\;.
\end{eqnarray}
We find that $\rho=1$. Naively, since $m^2_Z$ is modified to be $\bar m^2_Z = m^2_Z (1+z)$, this change seems should generate a non-zero value for the $T$ parameter and therefore $\rho \neq 1$. However, in our case, the neutral current also gets modified by a factor of 
$\sqrt{1+z}$. When taking the ratio for $\rho$, the factor $1+z$ cancels out.

We conclude that if processes involve only SM particles and only EM and weak interactions are probed without dark sector current included in the calculation (at tree or loop level), there is no physical effect due to a non-zero $\sigma$ showing up if dark photon is exactly massless.

Where can the kinetic mixing effect then be detected? Interaction with dark current must be involved in order to see any physical effect. 
Let us now discuss the case of search for dark photon that interacts with dark sector at tree level.

In general if there is dark current, the massless dark photon can act as a portal to probe dark sector particles.
This can be easily seen in case a) where the massless dark photon decouples from the SM but the photon can couple to dark current with a millicharge~\cite{Holdom:1985ag}. Thus the existence of millicharged particles can be signatures of the physical effect of massless dark photon that can be detected. There are many direct searches for millcharged particles  
at low-energy processes and high-energy processes~\cite{Prinz:1998ua,Gninenko:2018ter,Badertscher:2006fm,Magill:2018tbb,Singh:2018von,Davidson:2000hf, CMS:2012xi, Aubert:2008as,Haas:2014dda,Liu:2018jdi} yielding signatures of small ionization energy loss or missing energy, and also constraints on millicharged particles from astrophysics considerations~\cite{Vogel:2013raa}. Some of the constraints obtained are very stringent but depend on parameters in the dark sector particle properties generating the 
dark sector current $j_X^\mu$. Supposing that the dark current is generated by a Dirac dark fermion $f_\chi$ of the form $\bar j_X^{\mu}=\bar g_X\bar{f_\chi}\gamma^{\mu}f_\chi$, and the dark fermion has a millicharge  $Q_{\chi}=-\sigma c_{W}\bar g_{X}$. 
The constraints obtained from the above mentioned methods are usually applied to $\epsilon\equiv |Q_{\chi}/\bar{e}|$ as a function of the $f_\chi$ mass $m_{f_\chi}$. (Similar constraints for dark current generated by dark scalar particles can also be obtained.) For $10^{-4}\gev\lesssim m_{\chi}\lesssim 0.1\gev$, the parameter space of $\epsilon\lesssim 10^{-5}-10^{-3}$ has been excluded by the SLAC beam-dump experiment~\cite{Prinz:1998ua}. For a larger $m_{f_\chi}$ the collider experiments gives the bound $\epsilon\lesssim 0.2$~\cite{Davidson:2000hf}, which is expected to be much improved at the high-luminosity LHC~\cite{Haas:2014dda}, $\epsilon\lesssim 10^{-3}-10^{-2}$. The millicharged particle can couple to $Z$ boson with the coupling proportional to $-\epsilon \tan\theta_W$, thus assuming the decay $Z\to f_\chi\bar{f_\chi}$ for $m_{f_\chi}\lesssim 45\gev$~\cite{Davidson:1991si,Davidson:2000hf} one obtains that $\epsilon\lesssim 0.18$~\cite{Tanabashi:2018zz}.

\begin{figure}
\includegraphics[width=0.25\textwidth]{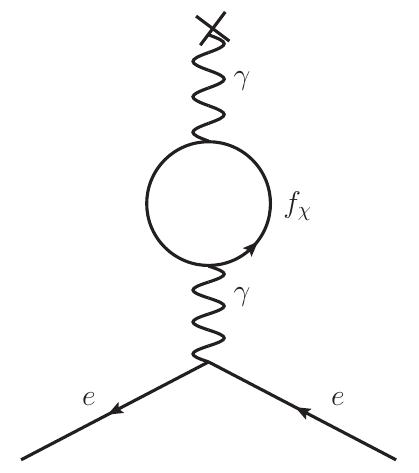}
\includegraphics[width=0.35\textwidth]{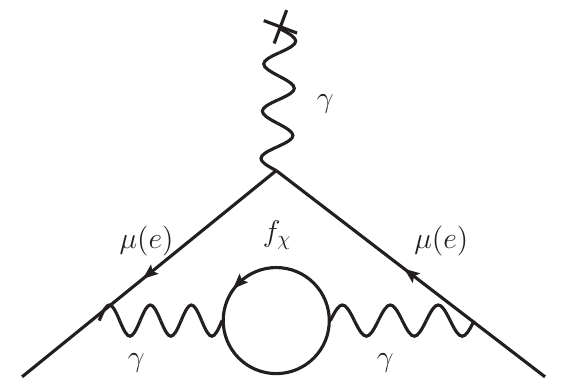}
\caption{Vacuum polarization diagrams contributing to and Lamb shift (left) and lepton $g-2$ (right) in the presence of millicharged particles. The cross vertex denotes the external source.}
\label{fig:g-2_2-loop}
\end{figure}

\begin{figure}
\includegraphics[width=0.4\textwidth]{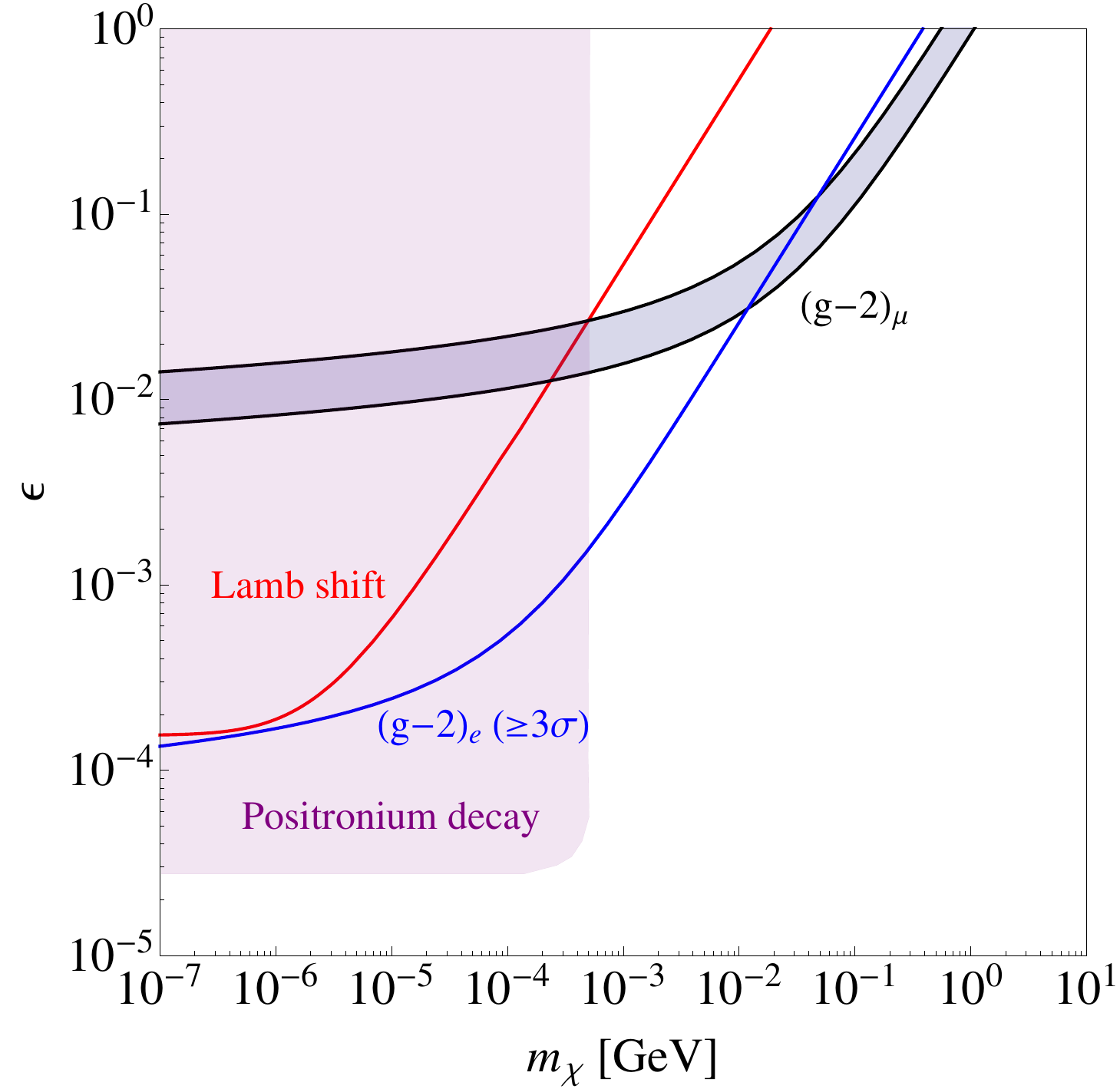}
\caption{Constraints on $\epsilon$ as a function of $m_{f_{\chi}}$ from the Lamb shift, electron and muon $g-2$ measurements and Positronium invisible decays. The black region is allowed by the muon $g-2$ measurements at 2-sigma level, while region above the red (blue) curve is excluded by the Lamb shift (electron $g-2$) measurement at 2-sigma (3-sigma) level. The purple region is excluded by the invisible decays of Positronium at 90\% C.L.~\cite{Badertscher:2006fm}.}
\label{fig:mchi_epsilon}
\end{figure}

The dark photon and dark current can also produce physical effects at loop level affecting some of the most precisely measured quantities, such as Lam shift and lepton $g-2$ through modifying the photon propagator~\cite{Dobroliubov:1989mr,Davidson:1991si}. Figure~\ref{fig:g-2_2-loop} (left panel) shows the corresponding vacuum polarization diagrams with millicharged particles in the loops. The contribution of Dirac fermionic millicharged particle to the Lamb shift between the levels of $2S_{1/2}$ and $2P_{1/2}$ of hydrogen atom was calculated in Ref.~\cite{Gluck:2007ia}, which is expressed as
\begin{align}
\delta E &= -\epsilon^2\dfrac{4\bar{\alpha}^3 m_e}{3\pi} \alpha^{*2} I(\alpha^*),\\
I(\alpha^*) &= \int_{0}^{1}du (1+\dfrac{u^2}{2})\dfrac{u \sqrt{1-u^2}}{(\alpha^*u+2)^4}
\end{align}
with $\alpha^*\equiv \bar{\alpha} m_e/m_{f_{\chi}}$ and $m_e$ being the  electron mass.

The contribution to the lepton $g-2$ due to the Dirac fermionic dark current, see right panel of Fig.~\ref{fig:g-2_2-loop}, can be obtained by scaling the QED mass-dependent correction to the lepton $g-2$ at the fourth order~\cite{Elend:1966vd,Passera:2004bj,Samuel:1990qf}, that is
\begin{eqnarray}
&&a_{\mu/e}^{\text{2-loop}}=\epsilon^2A_2({m_{f_\chi}\over m_{\mu/e}})\;,\;\;\;\;A_2(x) = {\alpha^2\over \pi^2} \int^1_0du\int^1_0dv {u^2(1-u)v^2(1-v^2/3)\over u^2(1-v^2)+4x^2(1-u)}\;.
\end{eqnarray}

Requiring that $a_{\mu}^{\text{2-loop}}$ satisfies $\Delta a_{\mu}=a_{\mu}^{\text{exp}}-a_{\mu}^{\text{SM}}= 268(63)(43)\times 10^{-11}$~\cite{Tanabashi:2018zz} and $\delta E$ is smaller than the experimental precision 0.02~MHz~\cite{Gluck:2007ia,Karshenboim:2005iy} at 2-sigma levels and $a_{e}^{\text{2-loop}}$ satisfies $\Delta a_{e}=a_{e}^{\text{exp}}-a_{e}^{\text{SM}}=-87\pm 36\times 10^{-14}$~\cite{Davoudiasl:2018fbb,Parker191,Aoyama:2017uqe} at 3-sigma level, we obtain constraints on $\epsilon$ as a function of $m_{f_{\chi}}$ shown in Fig.~\ref{fig:mchi_epsilon}. The black region is allowed by the muon $g-2$ measurements at 2-sigma level, while region above the red (blue) curve is excluded by the Lamb shift (electron $g-2$) measurement at 2-sigma (3-sigma) level. The purple region is excluded by the invisible decays of Positronium at 90\% C.L.~\cite{Badertscher:2006fm}. Interestingly, we find that the constraint from the electron $g-2$ is stronger than that from the Lamb shift. With future improvement in the sensitivity of electron $g-2$ measurement, one may gain more information about massless dark photon. Nevertheless, the constraints obtained in the same mass range for $m_{f_\chi}$ are weaker than that from SLAC beam-dump experiment~\cite{Prinz:1998ua}, invisible decays of Positronium~\cite{Badertscher:2006fm} and indirect constraint from the effective number of neutrino species bound~\cite{Vogel:2013raa,Brust:2013xpv}.

\begin{figure}
\includegraphics[width=0.35\textwidth]{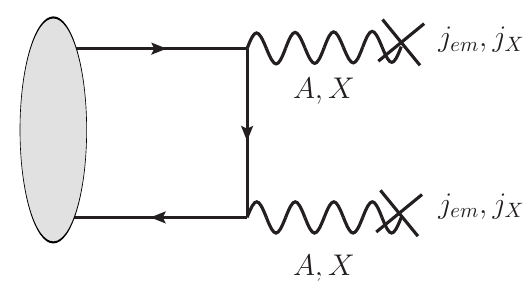}
\caption{Feynman digrams for the processes $h, pp\to AA,AX,XX$ connected to $j_{em}$ and $j_{X}$. The diagrams with the interchange of final gauge bosons are also involved.}
\label{fig:collider}
\end{figure}

Let us finally discuss the possibility of detecting dark photon effects without knowing details of dark current $j_X^\mu$.
 The processes involving two massless gauge bosons in the final states from the Higgs boson decay or proton-proton $(p p)$ collision are shown in Fig.~\ref{fig:collider}. To obtain the final results, one starts from $h, pp \to AA, AX,XX$ and analyze how $A$ and $X$ produce signal in the detectors. If $A$ or $X$ is detected by $j^\mu_{em}$, it is identified as a photon $(\gamma)$. The amplitude of $h,pp\to \gamma^*\gamma^*$ including the contributions from $A$ and $X$ is proportional to 
\begin{eqnarray}
e^4(2 R_A \times R_X + R_A\times R_A +  R_X \times R_X) = \left ({e^2\over 1-\sigma^2 c^2_W} \right )^2\;.
\end{eqnarray}
The total effect amounts to the redefinition $\bar e = e/\sqrt{1-\sigma^2c^2_W}$, just as in $g-2$ case, which is unobserved and independent of the angle $\beta$. Therefore, the rate at this order is equal to its SM value. If both $A$ and $X$ are on shell, effectively we can sum over the rates of $h,pp\to AA$, $AX$, $XX$, similar to that in the Compton scattering. This also leads to the SM rate into two real photons.

On the other hand, the amplitudes of $h, pp\to AA,AX,XX$ connected to $j_{em}j_{X}$ and $j_{X}^2$ depend on $\sigma$, which cannot be absorbed into the electric charge. It is easy to verify that the amplitudes are basis independent. The signature of $j_X^{\mu}$ that originates from dark sector particles escaping from detectors is missing energy. If  $A$ or $X$ coupled to $j_{X}^{\mu}$ result in off-shell dark photon, the amplitudes depend on dark charge $g_{X}$ and mass of dark particle. The detection rate is constrained by $\epsilon$, see Fig.~\ref{fig:mchi_epsilon}.

\section{ Conclusions}
\label{sec:summary}
In this work, we show that for the SM extended with a $U(1)_X$ gauge field having kinetic mixing with SM $U(1)_Y$ gauge field, the physical massless dark photon cannot be distinguished from the photon if re-writing gauge fields in the canonical form is the only requirement for removing the kinetic mixing term in the case with $m_X =0$. To make the points, we first show the details of two commonly used ways and show that they are related by an orthogonal transformation. Furthermore, one can arrive at a general mass eigenstate of photon and dark photon from case b) by an orthogonal transformation described by a rotation angle.   
We have shown that  such a mixing does not leave traces in the EM and weak interactions if only SM particles are involved. 
Physical effects of kinetic mixing will necessarily involve dark currents either in the form of missing energy (small ionization energy loss) or dark currents in loops. We have updated constraints on the millicharge from the Lamb shift and lepton $g-2$ measurements as a function of the millicharged particle mass.

\begin{acknowledgments}

We would like to thank Yi-Lei Tang and Fang Ye for fruitful discussions. XGH thanks R. Foot for discussions.
This work was supported in part by the NSFC (Grant Nos.11575115, 11735010 and 11975149), by Key Laboratory for Particle Physics,
Astrophysics and Cosmology, Ministry of Education, and Shanghai Key Laboratory for Particle
Physics and Cosmology (Grant No.15DZ2272100), and in part by the MOST (Grant No.MOST106-2112-M-002-003-MY3 ).
\end{acknowledgments}
\bibliographystyle{apsrev}
\bibliography{reference}

\end{document}